\documentclass[conference]{IEEEtran}
\IEEEoverridecommandlockouts
\usepackage{cite}
\usepackage{amsmath,amssymb,amsfonts}
\usepackage{algorithmic}
\usepackage{graphicx}
\usepackage{textcomp}
\usepackage{xcolor}
\usepackage{nomencl}
\usepackage{array}
\usepackage{url}
\usepackage{xurl}

\def\BibTeX{{\rm B\kern-.05em{\sc i\kern-.025em b}\kern-.08em
    T\kern-.1667em\lower.7ex\hbox{E}\kern-.125emX}}
\begin{document}

\title{The coordination between TSO and DSO in the context of energy transition - A review\\
\thanks{The authors would like to acknowledge the financial support for this work from NWO funded DEMOSES project (No. RS103940)}
}

\IEEEpubid{\makebox[\columnwidth]{979-8-3503-7973-0/24/\$31.00 ©2024 European Union\hfill} \hspace{\columnsep}\makebox[\columnwidth]{ }}

\author{
\IEEEauthorblockN{ Hang Nguyen, Koen Kok, Trung Thai Tran, Phuong H. Nguyen}
\IEEEauthorblockA{\textit{Department of Electrical Engineering, Electrical Energy Systems Group} \\
\textit{Eindhoven University of Technology}\\
Eindhoven, The Netherlands \\
\{t.h.nguyen, j.k.kok, t.t.tran, P.Nguyen.Hong\}@tue.nl}

}
\maketitle

\begin{abstract}
Nowadays, energy transition is ongoing in many countries, aiming to reduce dependence on fossil fuels and CO2 emissions. Besides the positive impacts on the environment, this transition brings technical challenges to the system operators, such as the intricacies of energy system integration, diminishing uncertainty, and incentivizing customers with advanced transaction models. The coordination between the Transmission system operator (TSO) and the Distribution system operator (DSO) is one of the most important aspects of encountering these obstacles. This coordination enhances the utilization of flexibility from Distributed energy resources (DERs) by incentivizing the market parties with better willingness to pay schemes. This paper provides an overview of the coordination schemes (CS), their classification, assessment of the current situation and the challenges associated with applying these schemes in practical context. The main purpose is to investigate the most effective way for TSO/DSOs to use the flexibility resource to maintain the balancing of the entire system while ensuring no congestion occurs in the network. A broad range of possible coordination schemes along with exploiting flexibility services is presented and the pros and cons are analyzed. Additionally, the study presents a general scenario that describes the interaction between the operators and the third party in providing service to the balancing market, considering cases with and without coordination.
\textbf{}
\end{abstract}


\section{Introduction}
Physically, the transmission network (TN) and distribution network (DN) are connected through transformers. However, they are operated separately by the transmission system operator (TSO) and the distribution system operator (DSO) due to differences in configuration and other technical characteristics \cite{26}. 
\begin{figure}[htbp]
\centerline{\includegraphics[scale=0.28]{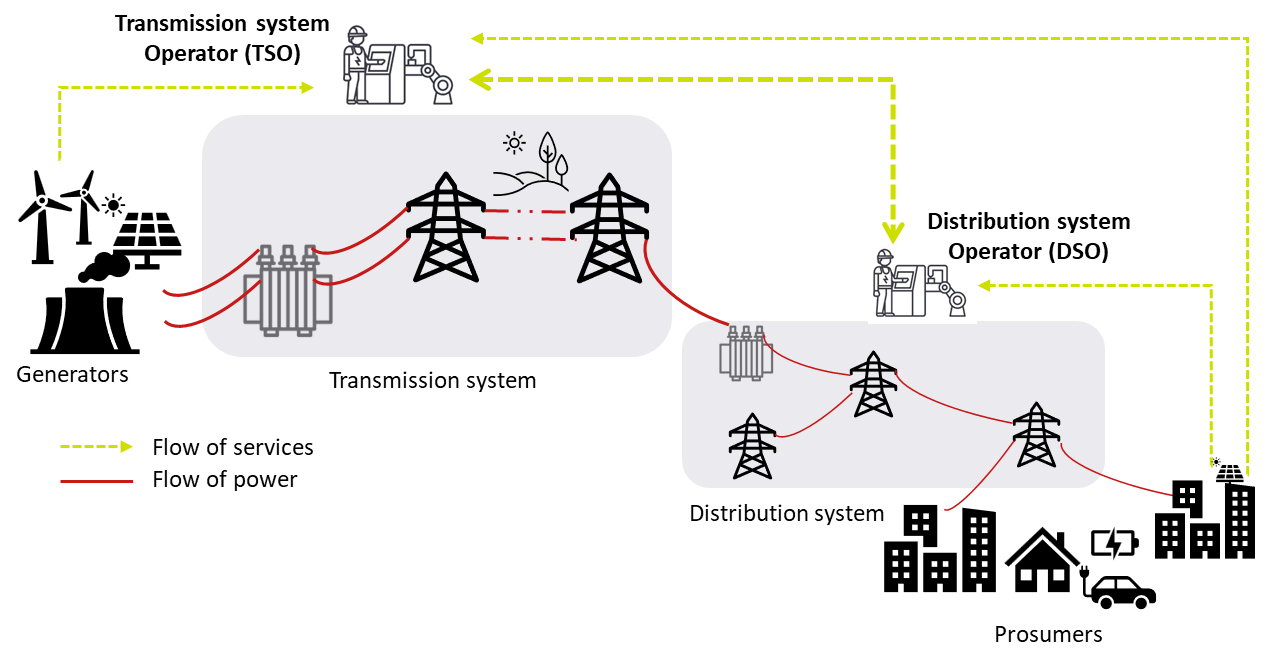}}
\caption{Interaction model between system operators and DERs.}
\label{fig}
\end{figure}
Based on the IRENA report \cite{1}, Fig.1 describes the interaction between TSO, DSO, service providers and the system through the flow of power and service. The flow of services is exchanged between TSO, DSOs, with or without the participation of third parties. In the current situation, TSO and DSO are responsible for planning, managing and expanding the TN and DN, respectively. Therefore, the rapid increase in Distributed energy resources (DERs) penetration to the DN has brought both opportunities and challenges to the systems. Using flexibility services provided by DERs helps to improve the overall efficiency and reliability of the electricity system\cite{1,25}. In return, this requires additional responsibility from the system operators to manage these flexible resources and ensure the system operates securely. 

According to Helena et al. \cite{5}, the coordination between TSO and DSO presents the interaction, roles and responsibilities of each system operator in procuring and using distribution grid-connected services.
Enhancing coordination between TSO and DSOs helps better use of the flexible resources from the distribution network, optimizes the overall operating costs and defers future investment in the network. Moreover, enhancing the coordination and data exchange between TSO and DSO is necessary for both long-term and operational planning, contributing to shared balancing responsibilities and grid monitoring capabilities of the system operators. 
This means increasing DSOs participation in the balancing process like the pre-qualification of the customer assets. On the TSO side, TSO can access the required data from DSO-connected grid users through DSO or indirectly from a neutral party to ensure the data integrity and without transferring the DSO metering responsibility \cite{13}. This coordination contributes to improving the system's efficiency and ensuring effective operation\cite{1}. Therefore, DSOs should have a more active role and responsibility in managing their local network\cite{4,3,21,28}. 

TSO-DSO coordination has been getting more attention from organizations and research groups in many countries. Various pilot projects in Europe have been initiated, such as SmartNet (Denmark, Italy and Spain)\cite{32}, CoordiNet (Greece, Spain and Sweden) \cite{34}, INTERRFACE (Italy, Bulgari, the Baltic and Nordic region countries) \cite{34,40}, InteGrid (European Commission) \cite{42}. Besides, there are several related projects like GOPACS (the Dutch TSO and its DSOs) \cite{37}, Enera (Germany), NODES (Norway and Germany), Piclo Flex (UK) \cite{41}, Soteria (Belgium), etc.\cite{4,36}.

CoordiNet and INTERRFACE\cite{22} look for a cost-efficient coordination scheme that can be scaled up and will be replicable across the EU energy system. Their goals include maximizing revenue streams, minimizing operational costs and increasing the share of renewable energy while ensuring the grid constraints. Additionally, the project SmarNet conducts comparisons across a wide range of coordination schemes (CS) to acquire ancillary services from the DN.
In GOPACS project, a trading platform that is able to resolve congestion issues in both TN and DN, was initiated by the Dutch grid operators. This platform connects not only the grid operators, but also the market parties and major consumers, ensuring the congestion issue in a certain part of the grid will not cause problems in other parts \cite{37}.
In Belgium, Soteria project focuses on optimizing the coordination to unlock unused distributed flexibility and considers demand response as a key enabler of energy transition.

These aforementioned projects share both common and distinct goals, statistic in \cite{31} reveal that there are five major coordination purposes, as voltage regulation \cite{23,24}, reactive power management \cite{10}, operational cost optimization \cite{22,34}, longterm operation planning \cite{27}, congestion management\cite{5}.
While TSO-DSO coordination is necessary and offers various benefits, it also faces certain obstacles.
One significant barrier is the inadequate information and communication technology infrastructure. Additionally, the majority of flexibility options reside with small-scale customers who cannot engage in the market, limiting the operators' ability to tap into this resource. Another limitation is the inability of DSO to access flexibility resources within their local networks across most European nations. Consequently, DSOs should be granted the capability to procure services that address issues within the distribution grid \cite{3}.
Similarly, the authors in \cite{34} emphasize the importance of empowering current and potential market actors, enhancing their role, and building trust by ensuring transparency. This CS is achieved via seamless information exchange and clear delineation of responsibilities.

Based on the previous systematic literature review from \cite{7,31}, the new international journals from ScienceDirect, IEEE Xplore database and international organization reports like ENTSO-E, IRENA, MIT CEEPR, and TENET’ document. The keywords used for the searches are TSO-DSO coordination, TSO-DSO, TSO/DSO, TSO-DSO integration, TSO-DSO interaction, coordination of transmission system operator and distribution system operator, and distributed flexibility. A review of TSO-DSO coordination is demonstrated in this paper to answer the following questions:
\begin{itemize}
\item What are the pros and cons of the coordination schemes?
\item What information should be exchanged between TSO and DSOs interface?
\item Which CS should be applied to suit the actual situation and effectively exploit the distributed resources?
\end{itemize}
The above questions will be discussed in the following parts. Section II presents the literature review on TSO-DSO coordination.  Section III discusses the pros and cons of the CSs. Next section presents the current situation and proposes a coordination scheme for the balancing market in the Netherlands. Finally, the conclusions are presented in section V.

\section{Literature review}
\subsection{The necessity of coordination and data exchange}\label{AA}
This section provides an overview of CSs, their necessity, different classifications and the physical pilots.

The authors in \cite{28} emphasize the importance of enhancing the interaction and coordination between TSO and DSO in both short-term action and long-term planning. 
The data exchange between TSO and DSOs should 
not only include power flow, forecasting data, and emergency situations as traditionally done, but also pricing, scheduling, and activation of their services.
Moreover, to prevent harmful interferences between the use of flexibility for balancing and congestion management, DSOs should receive relevant data to assess the impact of balancing services activation and determine the limits before activation\cite{13,39}. 

To underscore the necessity for coordination, Alejandro et al.\cite{25} show the potential cross impacts of uncoordinated activations. This shortage causes congestion or imbalance in the grid due to services activated in opposite directions and can jeopardize the benefits of using flexibility with the joined dispatch of the retailer. In \cite{36}, the authors demonstrate a TSO-DSO coordination in the case of UK, considering the high DER growth for future scenarios. The case studies give priority to DSO in using DERs to manage congestion in the distribution network, the remaining is sent to TSO by the traffic light mechanism. Results indicate that significant flexibility remains for the TSO even during periods of peak demand and maximum export. Furthermore, the objectives of TSO and DSOs are aligned in most cases. This mechanism is one solution to provide DERs flexibility to TSO while considering the distribution grid constraint. 

Talai Alazemi et al. \cite{31} conducted a systematic literature review based on the CSs, modeling or simulation tools, the corresponding optimization problem and their main objective. The findings underline the important role of DSO and TSO-DSO coordination. 
While research on this topic has received increased attention in recent years, there are still some gaps that need to be filled notably the absence of standardized communication solutions and the need for practical implementations to test and demonstrate feasible results. Moreover, there is a deficiency in a universal approach to solve all network management issues.

\subsection{Classification of coordination schemes}\label{AA}

Various frameworks have been introduced to facilitate the TSO-DSO coordination, each falling into distinct categories. Primarily, these frameworks can be differentiated based on market design, operational procedures, and the organization and exchange of data, categorizing them into centralized and decentralized models. The diagram in Fig. 2 explains the centralized approach in the left side and the decentralized approach in the right side. The black line shows the physical connection in the network, in which the flexible resource in high voltage (HV) level is connected to the transmission network, and the distributed energy resources (DERs) is connected to the medium and low voltage network (distribution network). In the centralized approach, TSO uses resources from both the transmission and distribution network (DERs) as illustrated by the green line. The red line represents the flow of control signal, TSO is responsible for managing the balancing, and congestion for the entire system. These huge tasks make this scheme less feasible to apply in reality.

According to statistics presented in [13], a significant portion of scholarly research leans towards the decentralized model,
accounting for more than 90\% of the research. Conversely, centralized models have attracted less attention.
This disparity is attributed to the high computational demands placed on the TSO and the necessity for intricate communication infrastructure in centralized systems.  


\begin{figure}[htbp]
\centerline{\includegraphics[scale=0.18]{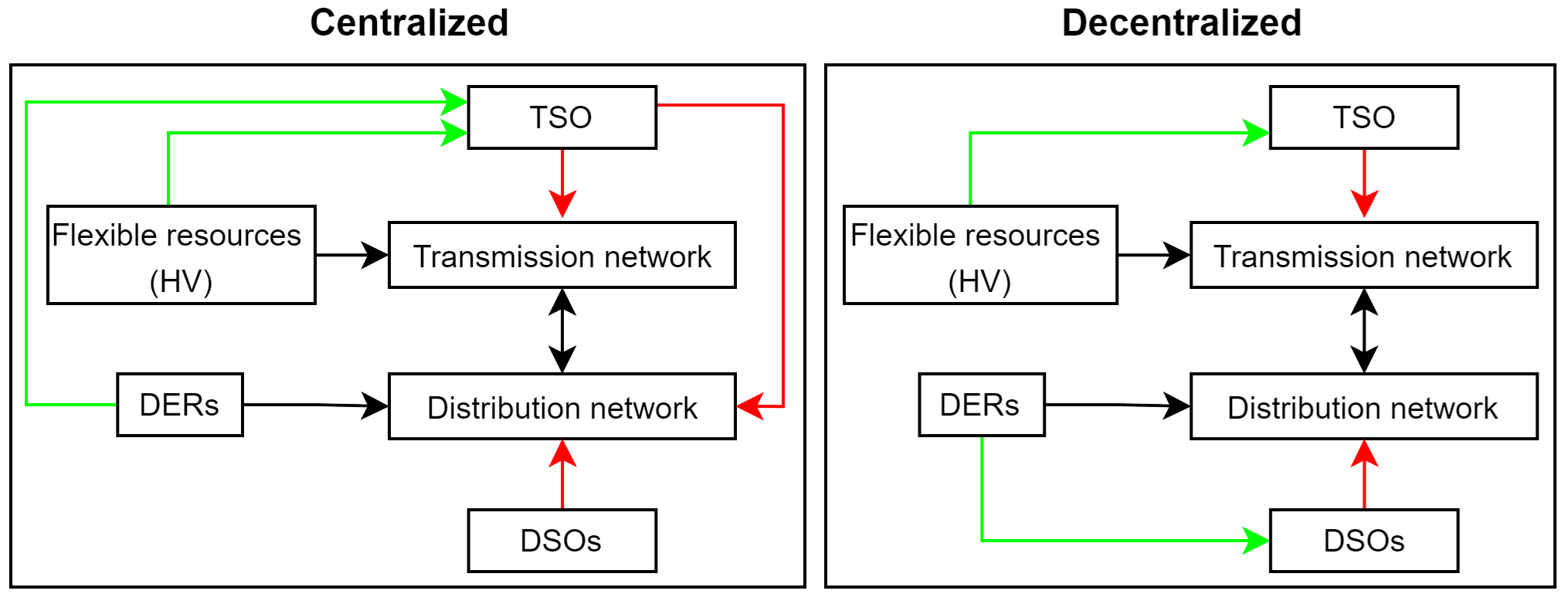}}
\caption{Centralized and decentralized approach.}
\label{fig}
\end{figure}


On the contrary, the decentralized approach equalizes the responsibility for both of them. TSO and DSOs use the RESs connected to their grid to manage congestion and maintain balance in their network.
The authors in \cite{30} the author compare the full smart grid (the aggregator and operators have full information about each other's operation) with the decentralized control (information about the grid is only partially available) and the scenario has no communication. Results show that decentralized control can enable around 90\% of the economic improvement of the fully controllable grid scenario thanks to the reduction of cost-intensive for upgrading the grid towards a smart grid. Similarly, in \cite{29} two decentralized schemes are compared with a benchmark centralized scheme. Their result shows that the first schemes bring the highest social welfare, while the decentralized way is more profitable for TSO when sharing the operational cost and reducing the burden of computational tasks. 


Arthur et al. \cite{8} categorize coordination models into three distinct schemes based on the validation process of DER bids and the entities responsible for this process. The first scheme is unique in avoiding any conflict of interest among DSOs. However, this model faces challenges as the volume of DERs increases, presenting a significant scalability issue for TSO. In contrast, the other two schemes empower DSOs with a more proactive role and increased responsibilities. Decision-makers are thus tasked with selecting the most suitable model, considering the current and future objectives of the system.
Likewise, the SmartNet report \cite{5,20} conducts its analysis by examining the relationships and roles between system operators, resulting in the classification of coordination models into five schemes. Among these, the common TSO-DSO ancillary market model aims to minimize total system costs. Meanwhile, the integrated flexibility market model prioritizes the allocation of flexibility to those with the highest willingness to pay with the involvement of an independent market operator.


\subsection{Physical pilots }\label{AA}
The majority of publications on TSO-DSO coordination are predominantly based on simulations. These simulations, while valuable, do not fully capture the complexities and realities of actual TSO-DSO coordination processes. Experiments conducted in hardware settings offer a more accurate reflection of the coordination dynamics and challenges.

The case study from Portugal \cite{9} highlights the implementation of an Inter Control Centre Protocol link between TSO and DSO. This system facilitates the exchange of critical real-time data, including power, reactive power, and the status of key equipment such as circuit breakers and switches. The improved coordination resulting from this setup enables the TSO to more accurately predict load changes. Simultaneously, it provides DSOs with the necessary insights to assess whether reconfiguration actions are required in response to significant failures within the transmission network

Currently, 4 of the 5 proposed schemes in the SmartNet project have been piloted in Italy, Denmark, and Spain scenarios \cite{20}. The result shows that the common marketplace should be implemented in the decentralized schemes. From the Danish scenario, the centralized schemes are the most effective and optimal ones if there is not a significant congestion issue in the DN. Besides, the cost of monitoring and controlling the DN could be higher than the over-investment cost at the beginning period while the ICT costs stay one order of magnitude lower than operational costs. Additionally, the reaction to commands from the operators in real-time seems to be too slow and further testing is needed to ensure responsiveness.


In the CoordiNet project \cite{36}, the researchers developed their coordination scheme drawing on insights from the SmartNet framework while introducing modifications through additional models: the multi-level, fragmented, and distributed market model. These models are designed to serve both TSO and DSO. Notably, in the fragmented market model, while services are available to both operators, TSO is restricted from utilizing DER flexibility.
From the Greek demo, both multi-level and fragmented CSs encouraged the formation of a local market that would consider in detail the components in the distribution grid and ensure secure and reliable operation of the distribution network.
The Sweden demo with the multi-level and distributed market model emphasizes the importance of putting DSO-TSO markets in the timeframes of the current spot market (day-ahead (DA), intraday market (ID)) without interference. The 2 CSs were selected based on the integration with the current market and regulation.  The result demonstrated that better coordination between DSO and TSO can give rise to more efficient grid use.

All the mentioned CSs in this project were shown to be suitable. However, there is no single approach adapted to the diverse TSO-DSO landscape. The selection depends on national characteristics, network topology, and the characteristics of the flexibility resources.

\section{Analytical comparison}
Multiple CSs have been proposed. Although these schemes are known by various names, they often share similar characteristics. Coordination between TSO and DSOs not only benefits the system operators in addressing network issues but also incentivizes the flexibility providers to engage in the market.
Flexibility has multiple dimensions like capacity, duration, ramp rate, direction, response time and location. Each type of flexibility should be used effectively in different dimensions. Therefore, the solution is to aggregate these resources together in a common flexibility market. In this common market, each participant uses service for different purposes, TSO use flexibility to ensure system balance, DSOs use flexibility to resolve local congestion, and third parties use flexibility to optimize their profit by correcting unbalanced positions in their portfolios. 

Each CS have a different impact on the system security base on how much, how fast and the characteristics of the service they can provide. However, in general, the more DERs resources available, the better the ability to ensure operation security. In turn, that requires more data exchange, making the system more vulnerable to external threats \cite{13,20,34}. The detail level of different kinds of data exchanges depends not only on the type of service but also on market rules.  Even though the SCADA system could support the SOs in the acquisition of relevant data from third parties, each SO should be responsible for its own IT system and develop the TSO-DSO interface to support data exchange in real time. Furthermore, it is essential to have an international data security standard and common communication framework for increasing harmonization of communication protocols and data exchange messages. In general, there are 6 main CSs. More detail about the characteristics, advantages, disadvantages, and DSO role of each coordination scheme is presented in Table I. Although each model has advantages and disadvantages, the presence of a common flexibility market, where all DERs from DN and large generators from the TN are provided to all the market parties like TSO, DSOs with or without third-party seems to be a promising approach to minimize the total operation cost and increase the social welfare.
\begin{table*}[htbp]
\caption{Pros and cons of the coordination schemes}
\fontsize{7}{8}\selectfont
\begin{center}

\begin{tabular}{| m{2.6cm}| m{5cm}| m{2.8cm}| m{3.2cm}|m{2.2cm}|}
\hline
\textbf{\textit{Coordination schemes}}& \textbf{\textit{Description}}& \textbf{\textit{Advantages}}& \textbf{\textit{Disadvantages}} & \textbf{\textit{DSO role}} \\
\hline

        TSO-managed model/ Centralize AS market model & 
        TSO contracts service directly with DERs provider, 
        TSO is the single buyer and performs economic dispatch of DERs&
        Simplify the coordination process.
        Minimize operational cost.
        No conflict of interest for DSOs.&
        TSO does not consider constraints in the DN due to lack of information.
        TSO faces huge computational 
        and modeling challenges. &
        DSO role is unchanged.\\
\hline

        DSO managed model/ Local AS market model & 
        DSOs aggregate and use DERs service for local congestion management, 
        then send the rest to the central market &
        DSO has a more important role and 
        consider local network constraints. &
        Conflict of interest among DSOs, 
        DSOs take huge computational and modeling challenges. &
        DSO has priority to use DER services.
        Manage congestion in the local network.\\
\hline        
        Shared balancing responsibility model/
        Fragmented market model &
        TSO and DSO manage their market.
        The responsibilities of the TSO and DSO are divided equally according to a predetermined schedule.&
        Respecting the DN constraints &
        TSO can not access the DERs resource. &
        Manage congestion and keep balancing the local system.\\
\hline
        TSO-DSO hybrid-managed model/ Multi-level market model &
        TSO and DSO manage their market.  TSO has access to DER service once DSOs validate service made by DERs to the central market. &
        Respecting the DN constraints. &
        DSOs take huge computational and modeling challenges.&
        Manage congestion and keep balancing the local system.\\
\hline 
        Common TSO-DSO AS market model &
        TSO and DSOs jointly manage to maintain balance and manage congestion for the entire system. TSO performs economic dispatch once DSOs validate DERs bids &
        Minimize total cost &
        Heavy mathematical for optimization and complex coordination processes.
        Conflict of interest among DSOs. &
        Local congestion management, cooperation with TSO when validating DERs bids. \\
\hline
        Integrated flexibility market model &
        TSO, DSO and commercial parties contract with 
        the independent market operators &
        Highest willingness to pay while 
        respecting  the DN constraints &
        Difficult to consider both TN and DN and the operational cost will not be optimized. &
        Manage congestion the local system.\\
\hline
\end{tabular}
\label{tab1}
\end{center}
\end{table*}
\section{Proposed coordination schemes}
\subsection{Sequence diagram}\label{AA}

\begin{figure}[htbp]
\centerline{\includegraphics[scale=0.148]{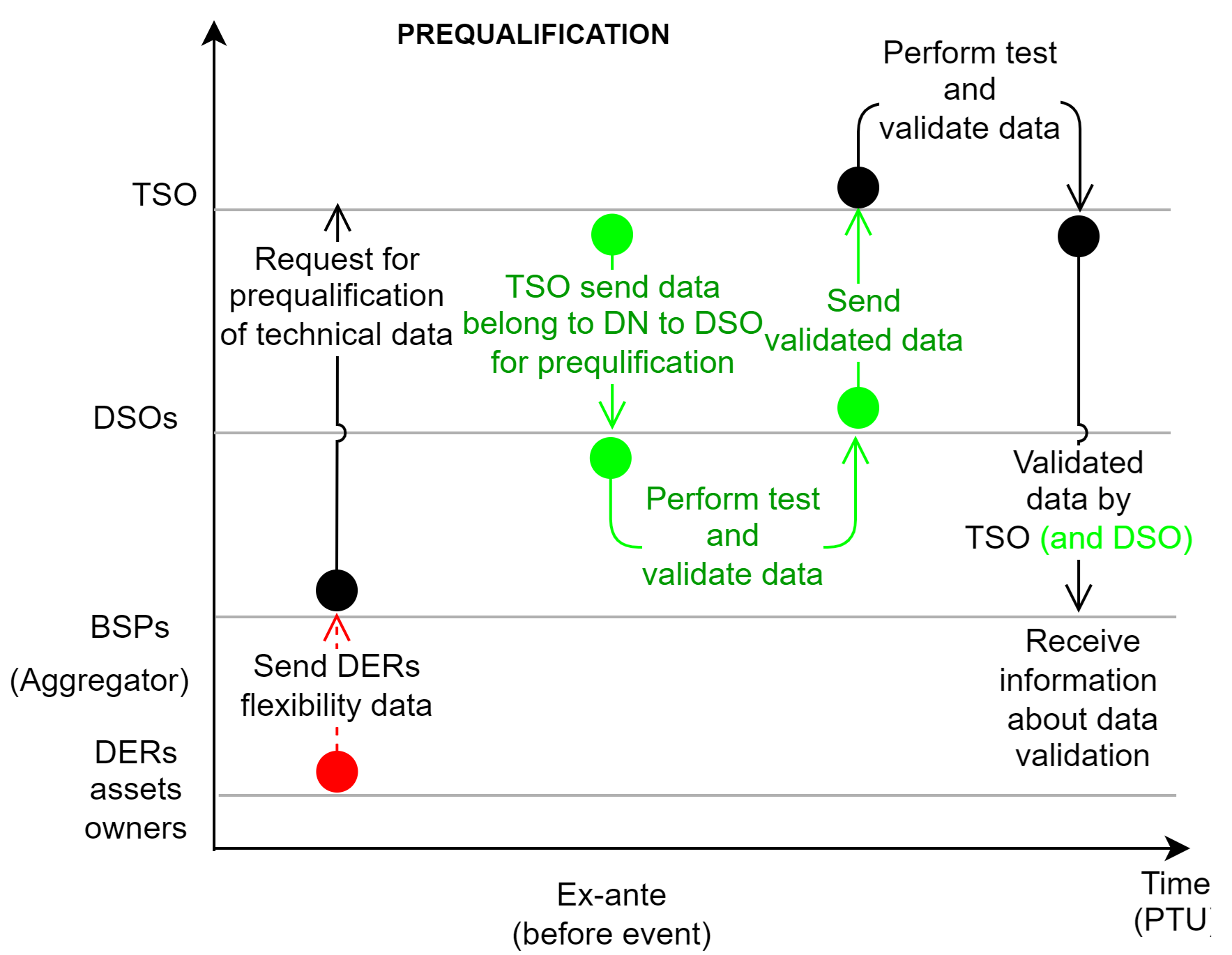}}
\caption{Sequence diagram of the prequalification process.}
\label{fig}
\end{figure}
Based on current situation of the electricity market in the Netherlands and the preceding discussion, a hybrid TSO-DSO coordination scheme is proposed and described in this section.
The sequence diagram in Fig. 3 presents an overview of the interaction between TSO, DSOs, and BSPs in the prequalification, offering and activation process, applied for the balancing market in the Netherlands based on the current situation and the scenario with the coordination. 
The black line presents the current situation, the green line presents the coordination between TSO and DSOs, the red line presents the flow of service from DERs. Besides, the dashed line shows that this interaction does not exist in reality.
This diagram describes the aFRR and mFRR services.
In which, BSPs take on the role of the aggregator who aggregates the DERs flexibility. Therefore, BSPs can provide balancing services from both large generators and DERs.
Before an asset belonging to a BSP can supply a balancing service (FRR) for TSO, the BSP needs to request the prequalification of the technical data to prove that the assets can deliver FRR under the specified technical minimum requirements\cite{38}. In this process, TSO can collaborate with DSOs by submitting the assets connected to the distribution grid under the operation of this DSO for prequalification. Once the DSOs finish the test and validate the technical data, they transmit the validated data to TSO. 
Subsequently, TSO performs tests and validates the assets connected to the transmission network. Finally, TSO sends the validated data to BSPs. 
BSP is required to retain this information with a minimum resolution of 4 seconds for 5 years or until re-qualification. 

The offering and activation process is presented in Fig.4. At the start of each imbalance settlement period (ISP), TSO aggregates the energy bids from BSPs to create the new merit order list (list of energy bids ordered by price). In this process, to alleviate the impact on the distribution network, TSO should coordinate with DSO by sending the MOL to DSO for validation, then DSOs perform power flow to check if any component in their network violates the constraint and clear the invalid bid or update the new value of the upper and lower boundary of the energy bid, then send the validated MOL to TSO. After that, TSO continues to verify until the MOL is accepted and sends the offer accepted confirmation to BSPs. However, there is no coordination between TSO and DSO in the offering process in reality. The aFRR is activated automatically by the load frequency control (LFC) following the delta setpoint provided by TSO. The offer accepted confirmation is the activation of the accepted bid for aFRR and 15 minutes before for mFRR. Based on the system's frequency deviation, TSO determines the volume that needs to be adjusted and requests BSPs to regulate upward or downward their offer following the delta setpoint. Tennet monitors the difference between the requested volumes and the total of BSP allocated every 5 minutes. TSO pays for BSP based on the balancing energy price and the total activation per ISP. 
In this process, BSPs have the right to activate the available assets that are most economically efficient for them. The utilization of flexibility for congestion management is not presented in this sequence diagram and is conducted in the DA and ID markets.
\begin{figure}[htbp]
\centerline{\includegraphics[scale=0.155]{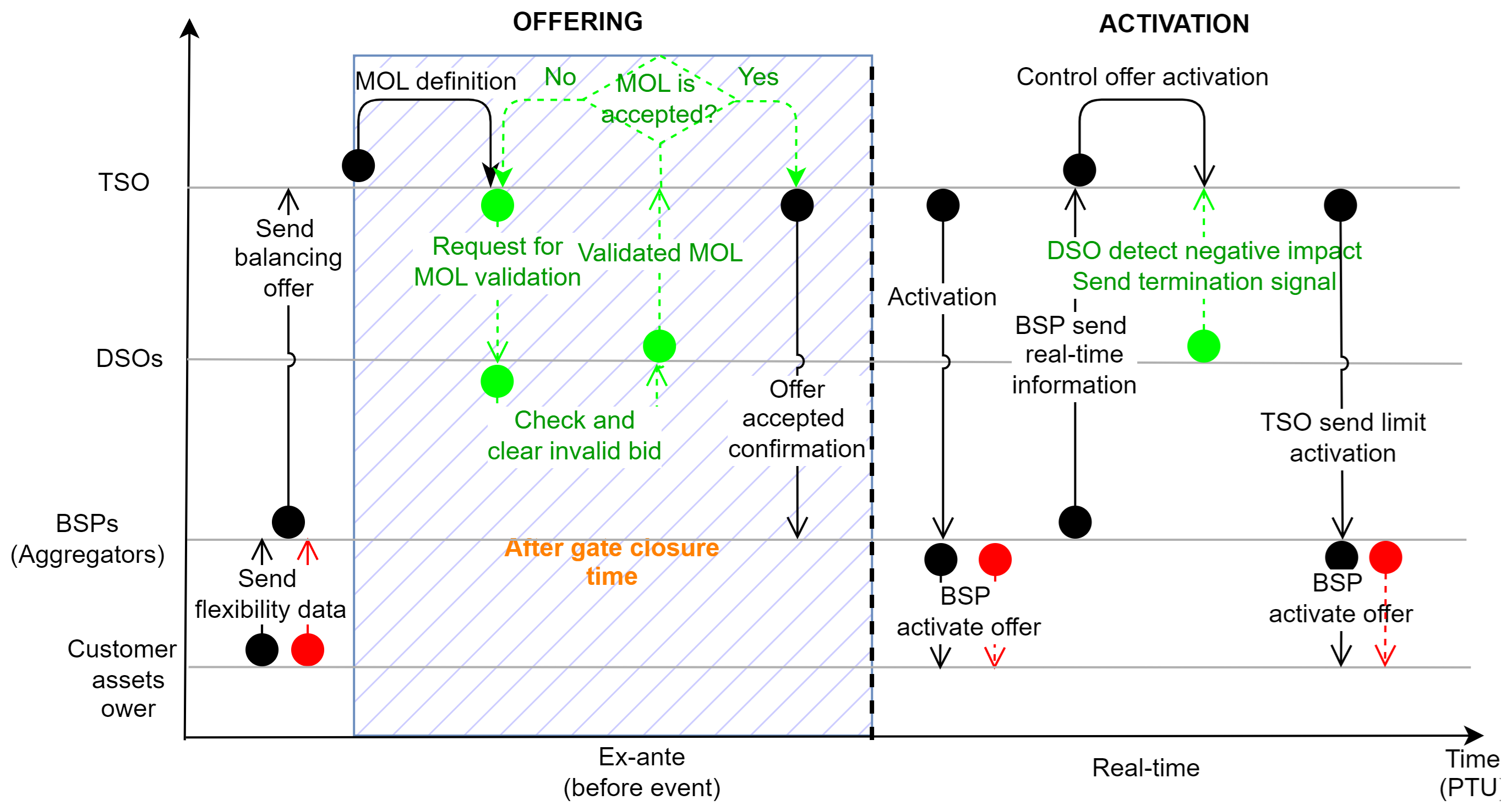}}
\caption{Sequence diagram of the offering and activation process.}
\label{fig}
\end{figure}
\subsection{Example}\label{AA}
Let's assume a BSP which have a stable production of 200MW can provide aFRR energy bid of upward and downward 20 MW. In which, 10 MW is aggregated from the DER flexibility. Besides, they have the lowest price compared with other BSPs and will be ordered in the first place on the merit order list. This BSP has a contract with TSO and submitted their bids to TSO before 2:45 pm the day prior and their assets were prequalified by TSO.
At 10:00, the system is short, TSO decides to activate the 10MW energy bid from the above BSP (In this case, the disturbance is not significant, if not, TSO has to activate all energy bids in parallel with the maximum ramping rate). The volume 200MW is considered as the reference value. TSO sends the delta setpoint to BSP with a maximum value of 1 MW and minimum time resolution of 4 seconds. The change of delta setpoint is larger than the ramp rate of the energy bids to make sure BSP assets have enough time to adjust their supply following this instruction. Therefore, from 10:01 to 10:06, every 30 seconds, TSO increases the delta setpoint to 1 MW until reaches 10 MW, and the setpoint will kept unchanged. However, the system is still short, this BSP can continue to increase till its maximum of 20MW to help the system balance as long as the average energy supply during this stable period is at least equal to the requested energy. The FVC will continuously monitor the volume of balancing energy that has been activated. BSP can tolerate their supply in the range [-10\%,20\%] of the setpoint. In case the BSP assets can not follow the setpoint, TSO has the right to withdraw the BSP qualification.
In the next ISP, this bid is deactivated due to its price or the activated bid is no longer available. TSO provides the setpoint to regulate back to the reference value of 200 MW.

However, activating this bid with a total volume of 20 MW causes congestion at the line behind the transformer at the coupling point to the distribution network because the DSO also activates the re-dispatch bid to resolve the congestion in another region in their network. In case TSO and DSOs have coordination in the offering process, DSO will perform power flow, update the limit and send the new boundary of 15MW to TSO. At the delivery time, the BSP can not supply over this limit and DSO will monitor and update the information to TSO. Therefore, with this interaction, the congestion will not occur anymore.
Therefore, with coordination, TSO can send their flexibility bid profile to DSOs, to make sure no flexibility bid from TSO will violate the constraint in the distribution network or be activated in the opposite direction with DSOs to alleviate the negative impact and reduce the total operational cost. The system operators work together to manage the common market and resolve the network issue.

\section{Conclusion}
To attain the zero-carbon target, the shift from fossil fuels to renewable energy is unavoidable and is being promoted by numerous nations and organizations. Nevertheless, there exists a trade-off between the environmental advantages and the challenges in energy system management. System operators must transform these challenges into opportunities by enhancing coordination among themselves in both short-term action and long-term planning.
This paper presents a review of the available TSO-DSO coordinations and provides insight into the CSs, type of data exchange, and the role of market participants for each scheme. Moreover, different schemes will have different orders of priority to use flexibility, the advantages and disadvantages.
It is important to highlight the necessity of the construction of a local flexibility market, where TSO, DSOs and other parties have access to the resources and DSOs have a more active role in using their resource.
A sequence diagram describes the interaction between TSO, DSO and BSP for the balancing market in the Netherlands based on the hybrid TSO-DSO managed model. There is coordination between TSO and DSO in the prequalification process. However, the shortage of real-time and near real-time coordination has the potential to impact the network. Future work will be based on the above assessment to quantify the issue in the transmission and distribution network considering the TSO-DSO coordination for both the current situation and projected scenarios. 
\bibliography{ref}
\end{document}